\documentclass[11pt,a4paper]{article}
\usepackage[latin1]{inputenc}
\usepackage{jheppub}
\usepackage{amsmath}
\usepackage{amsfonts}
\usepackage{amssymb}
\usepackage{graphicx} 
\usepackage{epsfig}
\usepackage{color}
\usepackage{amssymb}

\def\beq{\begin{equation}}
\def\eeq{\end{equation}}
\def\baq{\begin{eqnarray}}
\def\eaq{\end{eqnarray}}
\newcommand{\ee}[1]{\begin{equation}#1\end{equation}}
\newcommand{\ea}[1]{\begin{align}#1\end{align}}

\newcommand{\bk}{{\bf k}}

\providecommand{\f}[2]{\frac{{#1}}{{#2}}}

\title{Light scalars on cosmological backgrounds}

\author[a,b]{Tommi Markkanen}
\affiliation[a]{Department of Physics, Imperial College London, SW7 2AZ, UK}
\affiliation[b]{Department of Physics, King's College London, Strand, London WC2R 2LS, UK}

\abstract{We study the behaviour of a light quartically self-interacting scalar field $\phi$ on curved backgrounds that may be described with the cosmological equation state parameter $w$. At leading order in the non-perturbative 2PI expansion we find a general formula for the variance $\langle\hat{\phi}^2\rangle$ and show for several previously unexplored cases, including matter domination and kination, that the curvature of space can induce a significant excitation of the field. We discuss how the generation of a non-zero variance for $w\neq-1$ can be understood as a process of self-regulation of the infrared divergences very similarly to what is known to occur in de Sitter space. To conclude, the appearance of an effective mass due to self-interaction is generic for a light scalar in curved space and can have important implications for reheating, vacuum stability and dark matter generation.
}

\emailAdd{t.markkanen@imperial.ac.uk}
\emailAdd{tommi.markkanen@kcl.ac.uk}
\begin{document}
\begin{flushleft}
	\hfill		  IMPERIAL/TP/2017/TM/01
	\newline \hfill	\raggedleft		  KCL-PH-TH/2017-53
\end{flushleft}
\maketitle
\tableofcontents
\section{Introduction}
\label{sec:sum}
Scalar fields are an important ingredient of the Standard Model of particle physics and its extensions as well as crucial for Early Universe physics, in particular inflation. They are also the most studied example in the framework of quantum field theory in curved space which assumes the presence of a classical curved background \cite{Birrell:1982ix,ParkerToms}.

Most work investigating the behaviour a scalar field $\phi$ in curved space assume a free theory. An important exception is de Sitter space, which due to its simplicity and importance for the inflationary paradigm has been studied by many also for interacting theories. For some time it has been known that in de Sitter space the usual perturbative expansion for a massless self-interacting quantum scalar field is problematic due to secular terms that grow without bound \cite{Tsamis:2005hd,Kahya:2010xh} and furthermore infrared divergences in the propagators \cite{Seery:2010kh,Enqvist:2008kt}. These issues can however be overcome via non-perturbative resummation techniques \cite{Burgess:2009bs}. What these methods reveal is that in de Sitter space interactions lead to the generation a non-zero variance $\langle\hat{\phi}^2\rangle$ giving rise to an effective mass, which will regulate the infrared behaviour of the theory \cite{Starobinsky:1994bd,Garbrecht:2014dca,Garbrecht:2013coa,Garbrecht:2011gu,Serreau:2011fu}. 

The stochastic framework \cite{Starobinsky:1986fx,Starobinsky:1994bd} is the most widely used approach for discussing scalar fields on de Sitter backgrounds \cite{Prokopec:2011ms,Hardwick:2017fjo,Vennin:2016wnk,Vennin:2015hra,Nurmi:2015ema,Moss:2016uix,Kainulainen:2016vzv,Prokopec:2007ak,Assadullahi:2016gkk,Nakao:1988yi,Nambu:1988je,Linde:1993xx,Karakaya:2017evp,Finelli:2010sh,Finelli:2008zg}, but they may also be approached via the 2-particle-irreducible (2PI) functional techniques \cite{Berges:2004yj,Blaizot:2001nr,Cornwall:1974vz} and there are already several works investigating scalar fields in de Sitter space via 2PI and related approaches \cite{Sloth:2006az,Herranen:2013raa,Serreau:2011fu,Riotto:2008mv,Ramsey:1997qc,Tranberg:2008ae,Serreau:2013psa,Ziaeepour:2017wdp,Arai:2011dd,Arai:2012sh,Nacir:2013xca,Prokopec:2011ms}. Furthermore, recently in \cite{Gautier:2013aoa} de Sitter space was approached via the Schwinger-Dyson equations and in \cite{Serreau:2013eoa,Guilleux:2015pma,Guilleux:2016oqv} nonperturbative renormalization group techniques were used.

Similar infrared divergences as encountered in de Sitter space have been known to arise also on other homogeneous and isotropic backgrounds \cite{Ford:1977in}, but in general the de Sitter solution has received the most attention, however see the recent works \cite{Lankinen:2016ile,Lankinen:2017rvi,Higuchi:2017sgj} analysing related issues beyond de Sitter space. In particular, interacting theories as studied here are currently largely unexplored. 

The object of this article is to provide solutions for light self-interacting scalar fields on cosmologically relevant backgrounds beyond de Sitter space. We will especially focus on the possible generation of a non-zero variance. Since the generalization of the stochastic approach beyond de Sitter is non-trivial -- although an interesting question in its own right, see for example the recent works \cite{Glavan:2013mra,Cho:2015pwa,Prokopec:2015owa,Glavan:2015cut,Glavan:2017jye} -- here we make use of the leading order truncation in a 2PI expansion. As we will show, it lends itself also to other homogeneous and isotropic spaces.

As our model we choose a generic decoupled sector consisting of a light self-interacting scalar singlet with the action
\ee{S=-\int d^4x\,\sqrt{|g|}\bigg[\f{1}{2}(\nabla\phi)^2+\f{1}{2}m^2\phi^2+\f{\xi}{2}R\phi^2+\f{\lambda}{4}\phi^4\bigg]\,,\label{eq:act}} 
on a background that is assumed to be of the Friedmann--Lema\^{i}tre--Robertson--Walker (FLRW) form, with the line-element given in cosmic time as
\ee{ds^2=-dt^2+a(t)^2d\mathbf{x}^2\,.\label{eq:le}}
In our analysis the notion of a light field is defined as a field that has a negligible mass that can be ignored with respect to the background Hubble rate and restricts the validity of our results to apply only in the case where $m/H\ll 1$ holds, with $H\equiv \dot{a}(t)/a(t)$. This poses a limit on the mass parameter present at tree-level as well as the running mass generated by quantum loops, which in the not completely decoupled case also results in restrictions on other sectors. However, no restrictions are imposed on the non-minimal coupling between the field and the scalar curvature of space $R$ as mediated by $\xi$. Following the standard cosmological framework, we will furthermore assume that the background may be characterized via a constant equation of state parameter $w$ relating the energy and pressure densities, $\rho$ and $p$, in the parametrization (\ref{eq:le}) as
\ee{p=w\rho\,;\qquad \rho = T_{00}\,,~~p = T_{ii}/a(t)^2\,,}
where $T_{\mu\nu}$ is the energy-momentum tensor of the background.

We will throughout leave the $t$-dependence in the scale factor implicit, $a(t)\equiv a$, use natural units with $c\equiv\hbar\equiv 1$ and 
follow the $(+,+,+)$ conventions of \cite{Misner:1974qy} for Einsteinian gravity.

\section{Free scalar field on a background with $p=w\rho$}
\label{sec:mod}
Setting $\lambda=0$, the action (\ref{eq:act}) gives rise to the equation of motion
\ee{\left(-\Box+m^2+\xi R\right)\hat{\phi}=0\,, \label{eq:eom}}
whose solutions, as usual, can be expressed as a mode expansion. Using conformal time defined as 
\ee{\eta=\int^{t}\f{dt'}{a(t')}\,,\label{eq:conft}}
the line element becomes
\ee{ds^2=a^2(-d\eta^2+d\mathbf{x}^2)\,,}
with which the properly normalized modes read
\ee{\hat{\phi}=\int \f{d^{3}\mathbf{k}\, e^{i\mathbf{k\cdot\mathbf{x}}}}{\sqrt{(2\pi )^{3}a^2}}\left[\hat{a}_\mathbf{k}^{\phantom{\dagger}}f^{\phantom{\dagger}}_{k}(\eta)+\hat{a}_{-\mathbf{k}}^\dagger f^*_k(\eta)\right]\,\label{eq:adsol2}\, ,}
with $[\hat{a}_{\mathbf{k}}^{\phantom{\dagger}},\hat{a}_{\mathbf{k}'}^\dagger]=\delta^{(3)}(\mathbf{k}-\mathbf{k}'),~~[\hat{a}_{\mathbf{k}}^{\phantom{\dagger}},\hat{a}_{\mathbf{k}'}^{\phantom{\dagger}}]=[\hat{a}_{\mathbf{k}}^{{\dagger}},\hat{a}_{\mathbf{k}'}^\dagger]=0$, 
where $\mathbf{k}$ is the co-moving momentum and $k\equiv|\mathbf{k}|$. From (\ref{eq:eom}) and (\ref{eq:adsol2}) one gets the mode equation
\beq
f''_{k}(\eta)+ \bigg[\bk^2 + a^2 m^2+ a^2 \left(\xi-\frac{1}{6}\right) R \bigg] f_{k}(\eta) = 0 \,,
\label{eq:mode}\eeq
where the primes denote derivatives with respect to conformal time.
Solving equation (\ref{eq:mode}) for the case of inflation with $w=-1$, or more precisely de Sitter space, is very well known from the cosmological context. However, much less work exists for backgrounds with $w\neq-1$, which we will address next. 

It is a straightforward exercise to use the Friedmann equations 
\ea{
\begin{cases}\phantom{-(}3H^2M_{\rm P}^2&= \rho\\ -(3H^2+2\dot{H})M_{\rm P}^2 &= p=w\rho \end{cases}\,,\label{eq:e}}
where $M_{\rm P}^2\equiv (8\pi G)^{-1}$ and show that the scale factor and Hubble rate can be written 
as
\ee{a=\bigg(\f{t}{t_0}\bigg)^{\f{2}{3(w+1)}}\,,\quad H=\f{2}{3(w+1)t}\,.\label{eq:rehaH}}
For our purposes it is more convenient to use conformal time (\ref{eq:conft}), where one has
\ee{a=\bigg(\f{\eta}{\eta_0}\bigg)^{\f{2}{3w+1}}\,,\quad \eta=\f{2}{3w+1}\f{1}{aH}\,.\label{eq:rehaH2}}
Note that $\eta$ changes sign at $w=-1/3$, but from (\ref{eq:rehaH}) we also see that this point separates a decreasing and increasing $(aH)^{-1}$, so for all $w$ one has $d\eta/dt>0$. Finally, we can write the scalar curvature as
\ee{R=6\f{a''}{a^3}=12\f{1-3w}{(3w+1)^2}(a\eta)^{-2}=3(1-3w)H^2\,.\label{eq:R} }

From now on we will specialize to the case of a light field by neglecting the mass parameter and with the help of (\ref{eq:rehaH2}) and (\ref{eq:R}) the mode equation (\ref{eq:mode}) assumes the form of Bessel's equation, i.e we get
\beq
f''_{k}(\eta)+ \bigg[\bk^2 - \f{\beta^2-1/4}{\eta^2}\bigg]f(\eta) = 0 \,,
\label{eq:modeireh}\eeq
with the definition
\ee{\beta^2=\f{1}{4}+12\bigg(\f{1}{6}-\xi\bigg)\f{1-3w}{(3w+1)^2}\,.\label{eq:nu2}}
An important feature worth emphasizing is that (\ref{eq:modeireh}) with (\ref{eq:nu2}) has a finite limit at $w=-1$, which may seem surprising since obviously the solutions (\ref{eq:rehaH}) are ill-defined for this choice\footnote{This feature is in fact only an artifact of the usual parametrizations of the solutions used in (\ref{eq:rehaH}). We can rectify this by instead parametrizing the solutions to (\ref{eq:e}) as \ee{a=\bigg[1+\f{3(w+1)H_0t}{2}\bigg]^{\f{2}{3(w+1)}}\,;\quad H=\f{H_0}{1+\f{3(w+1)H_0t}{2}}\,,\label{eq:rehaH3}}
where $H_0=H(0)$. These forms have the usual limit at $w\rightarrow-1$ and are physically equivalent to (\ref{eq:rehaH}) and hence also lead to (\ref{eq:modeireh}).}. There is however a divergence at the point $w=-1/3$, but as we will show in section \ref{sec:res} it disappears in the result for $\langle\hat{\phi}^2\rangle$, indicating that it has no physical significance. This is expected on the basis that the scalar curvature $R$ is well-behaved at $w=-1/3$.
\subsection{The generalized Bunch-Davies vacuum}
The generic solution for (\ref{eq:modeireh}) is a linear combination of the two Hankel functions
\ee{f_k(\eta)=C_1(k)\f{1}{2}\sqrt{-\pi\eta}H^{(1)}_\beta(-k\eta)+C_2(k)\f{1}{2}\sqrt{-\pi\eta}H^{(2)}_\beta(-k\eta)\,.\label{eq:modeireh1}}
For de Sitter space, $w=-1$, the standard choice for the vacuum is to choose $C_1(k)=1$ and $C_2(k)=0$ i.e. the Bunch-Davies vacuum \cite{Chernikov:1968zm,BD}. 
Formally, one may obtain the Bunch-Davies vacuum with the requirement that at early times, $t\rightarrow-\infty$, the behaviour of all $f_k(\eta)$ is independent of the curvature of space i.e. when $|k\eta|\rightarrow\infty$ we require our mode to behave as a positive frequency plane wave 
\ee{f_k(\eta)
\sim\f{e^{-ik\eta }}{\sqrt{2k}}\,,\label{eq:confe}}
which can be obtained from (\ref{eq:modeireh1}) by making use of the asymptotic form for the first Hankel function, $H_\nu^{(1)}(x) \sim \sqrt{{2}/({\pi x})}\exp\left(ix\right)$, for large positive $x$ with positive $\nu$. Note that $"\sim"$ signifies factors of modulus unity. 
The Bunch-Davies vacuum also has the appealing feature that it is a late time attractor: if we impose that (\ref{eq:confe}) is satisfied for modes of any solution in the high ultraviolet (UV), $k\rightarrow\infty$, this translates as requiring the UV portion of all solutions to coincide with the Bunch-Davies vacuum. But since in de Sitter $-\eta\rightarrow\infty$ when $t\rightarrow-\infty$, at late times all modes even ones deep in the infrared (IR) were at early times in the UV with $|k\eta|\rightarrow\infty$ and hence must coincide with the Bunch-Davies vacuum. For more explanation, see for example \cite{Markkanen:2016aes}. 

For $w\neq-1$ we will use the same requirement of correspondence with the positive frequency plane wave at $|k\eta|\rightarrow\infty$ to motivate our choice of vacuum. As discussed below (\ref{eq:rehaH2}) for $w<-1/3$ conformal time $-\eta$ approaches 0 for large cosmic time $t$ and hence our requirement translates to demanding all modes to coincide with (\ref{eq:confe}) at $t\rightarrow-\infty$, very much like for the Bunch-Davies vacuum in de Sitter space. But for $w>-1/3$ the opposite is true or that we require the modes to match with (\ref{eq:confe}) at $t\rightarrow+\infty$ . 
This leads us to the following choice of vacuum
\ee{f_k(\eta)=\f{1}{2}\sqrt{-\pi\eta}H^{(1)}_\beta(-k\eta)\,,\label{eq:rehmode}}
where for a positive index $\beta$ the negative arguments may be obtained via $H^{(1)}_\nu (-|x|)=-e^{-i\nu\pi}H^{(2)}_\nu (|x|)$ for which the large argument asymptotics come with the help of $H_\nu^{(2)}(x) \sim \sqrt{{2}/({\pi x})}\exp\left(-ix\right) $.

We emphasize that our choice possesses several natural features: for $w<-1/3$ due to the already mentioned behaviour of $-\eta$ being a decreasing function of time for $w<-1/3$, (\ref{eq:rehmode}) exhibits the same attractor nature as the Bunch-Davies vacuum does in de Sitter space. For $w>-1/3$ at late times the co-moving horizon radius $\propto \eta$ can be seen from (\ref{eq:rehaH2}) to approach infinity which for any local observer indicates a spacetime indistinguishable from flat space, and is reflected by the solution (\ref{eq:rehmode}) by coinciding with the positive frequency plane wave (\ref{eq:confe}) i.e. the standard Poincar\'e invariant Minkowski vacuum.

In the situation where the equation of state parameter is a constant for only a finite duration after which evolving in some unspecified manner the choice (\ref{eq:rehmode}) for cases with $w>-1/3$ becomes less motivated as demanding the late time evolution to coincide with a plane wave would no longer constrain the finite intermediate range with a constant $w$. Here there hence is no natural unique candidate as the choice of vacuum in contrast to $w<-1/3$ implying a sensitivity on the initial conditions appropriate for the particular problem one is studying. However, in a case paralleling the current cosmological paradigm where the Universe was first for a very long time characterized by $w=-1$ and then sufficiently slowly evolved into some $w>-1/3$ the choice (\ref{eq:rehmode}) again seems motivated. 

Finally, at the limit of a conformal theory with $\xi=1/6$ or a background with a conformal equation of state $w=1/3$ as given by radiation, scales such as $H$ should not be visible in the solutions and indeed we have from (\ref{eq:nu2}) $\beta=1/2$ and again correspondence with (\ref{eq:confe})\footnote{\ee{H^{(1)}_{1/2}(x)\sim \sqrt{\f{2}{\pi x}}e^{i x}
\,.}}. 

Trivially, for $w=-1$ the choice (\ref{eq:rehmode}) coincides with the Bunch-Davies vacuum. We will name (\ref{eq:rehmode}) as the \textit{generalized Bunch-Davies vacuum} even though in other works exploring similar issues in spacetimes with  $w\neq-1$ the word 'generalized' is left out. The reason for our nomenclature is that invariance under the symmetries of de Sitter space is often used as a defining feature for the Bunch-Davies vacuum and for spacetimes with $w\neq-1$, in particular when the evolution is not close to de Sitter, this definition cannot be invoked.
\subsection{Infrared divergences}
For an inflating background it is well-known that a traditional loop expansion is problematic due to an IR amplification of quantum modes, not much unlike what is encountered in thermal field theory \cite{Kapusta:2006pm}. This can be seen by calculating the leading IR limit of the variance for a free field\footnote{See equation (\ref{eq:asymp}) for the small argument asymptotic form of $|H^{(1)}_\nu(x)|$.}
\ee{\langle\hat{\phi}^2\rangle
\propto H^2\int^{\Lambda_{\rm IR}}_0dx\,x^{2-2\beta}\,.\label{eq:freev}}
For $w=-1$ with a massless and minimally coupled field one has $\beta\rightarrow3/2$ and the above gives divergence at $x\rightarrow0$. This divergence is a manifestation of the fact that a massless non-interacting scalar field is incompatible with de Sitter invariance \cite{Allen:1985ux}. However, the self-interacting massless theory in de Sitter is perfectly well-behaved, but in order to see this one must resort to more involved approaches than the standard (perturbative) loop expansion. 

As discussed in the introduction, nonperturbative approaches are required in order to form well-behaved expansions for an interacting theory in de Sitter space. By far the most popular and calculationally convenient is the stochastic approach. 
However, as the stochastic method makes use of features that are specific to de Sitter space it is not obvious how one should generalize it when studying backgrounds with $w\neq-1$, although, see the works \cite{Glavan:2013mra,Cho:2015pwa,Prokopec:2015owa,Glavan:2015cut,Glavan:2017jye} addressing this very issue. From (\ref{eq:freev}) and (\ref{eq:nu2}) we can furthermore see that for many combinations of $\xi$ and $w$ one may obtain $\beta\geq3/2$ leading to an IR divergence and indicating the need to go beyond standard perturbation theory. Fortunately, as we will show the 2PI approach is perfectly applicable for backgrounds with $w\neq-1$.


\section{Non-perturbative resummation}
This section introduces only the bare essentials of the 2PI technique as required by our calculation omitting many key details and concepts. For more information we refer the reader to \cite{Berges:2004yj}.

\begin{figure}
\begin{center}
\includegraphics[width=0.9\textwidth,angle=0,origin=c,trim={3.9cm 13.8cm 4cm 13.8cm},clip]{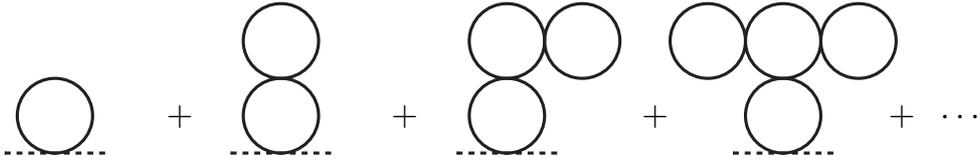}
\end{center}
\caption{\label{fig:2pi}The 2PI Hartree approximation. The sum in the above is included in the solution for the mode via (\ref{eq:modehar}) and is to be understood to consist of all graphs that can be formed by successive insertions of a "tadpole"-loop.}
\end{figure}

The 2PI technique is an inherently non-perturbative approach to field theory that relies on choosing particular topological classes of Feynman diagrams, usually containing an infinite number of graphs, and then includes the contribution of all such diagrams in order to obtain the "resummed" propagator. More or less the only analytically tractable 2PI approximation is the Hartree approximation \cite{Cooper:1994hr,Cooper:1996ii}, which consists of summing all "super-daisies" that can be constructed with the simple tadpole diagram as illustrated in Fig. \ref{fig:2pi}. This results from solving the mode equation supplemented with the one-loop effective mass from interactions, which for our theory reads
\beq
F''_{k}(\eta)+ \bigg\{\bk^2 + a^2 \bigg[m^2_0+\left(\xi_0-\frac{1}{6}\right) R +3\lambda_0\langle\hat{\phi}^2\rangle_0\bigg]\bigg\} F_{k}(\eta) = 0 \,,
\label{eq:modehar}\eeq 
where the subscripts of "0" indicate that we are dealing with unrenormalized quantities, in particular $m^2_0$, $\xi_0$ and $\lambda_0$  contain the necessary counter terms and the renormalized mass will be set to zero. Here the mode is denoted with $F_k(\eta)$ to emphasize that unlike the $f_k(\eta)$ in section \ref{sec:mod} it represents the solution to infinite order in $\lambda$. We will assume throughout that the 1-point function vanishes as an initial condition, $\langle\hat{\phi}\rangle=0$. 

The key idea behind this approach is that when solving equation (\ref{eq:modehar}) at no point should an expansion in terms of $\lambda$ be used and in particular $\langle\hat{\phi}^2\rangle_0$ denotes the full interacting variance for which (\ref{eq:modehar}) implicitly defines a self-consistent solution. In practice, more or less the only analytic way of obtaining a solution is via a conveniently chosen ansatz. 

In de Sitter space when only local diagrams are included as in the Hartree approximation of Fig. \ref{fig:2pi} one may require that the full propagator merely generates a constant effective mass contribution based on de Sitter invariance, and indeed this assumption allows one to find a solution for (\ref{eq:modehar}), see for example \cite{Herranen:2013raa}. For $w\neq-1$ when the Hubble rate is no longer a constant this assumption seems less motivated. On the other hand, when the tree-level mass of the field is assumed to be small the only scale that could potentially appear in the mode equation is $H$. This leads us to write an ansatz of the form
\ee{m^2_0+\xi_0R+3\lambda_0\langle\hat{\phi}^2\rangle_0 = \theta R\,,\label{eq:m2pi}}
where the $\theta$ is a constant that is to be self-consistently solved by using (\ref{eq:m2pi}) in (\ref{eq:modehar}). In the 2PI framework an equation of the type (\ref{eq:m2pi}) that relates an effective mass of a mode to the variance and hence implicitly to itself is commonly referred to as the gap equation. 

If one uses (\ref{eq:modehar}) with (\ref{eq:m2pi}) for studying theories that are not strictly massless and decoupled, the main constraint implied by (\ref{eq:m2pi}) is that any mass terms present at tree-level (and potentially amplified by loops from self-interactions) or couplings to other massive fields are small enough so that they may be ignored. Importantly, this condition must apply throughout the time scales one is considering. The longer the time the more stringent this condition is since generically when $w\neq-1$, $R$ is a decreasing function as $R\propto t^{-2}$. After a sufficiently long time a mass term will unavoidably become relevant resulting in a breakdown of the assumption (\ref{eq:m2pi}).

Finally, we draw attention to an important issue concerning the limitations of the Hartree approximation. It was already famously discussed in \cite{Starobinsky:1994bd} that it captures only the Gaussian part of the variance while the stochastic framework correctly includes also the non-Gaussian features of the probability distribution function of a theory with a quartic self-interaction. Comparing the results of \cite{Prokopec:2015owa,Cho:2015pwa} to \cite{Herranen:2013raa} one may see that this discrepancy persists beyond de Sitter and is thus present also in our solutions. However, if one is only interested in the magnitude of the variance and not in higher order correlators the numerical agreement between the two approaches is quite good \cite{Starobinsky:1994bd}.

\subsection{Calculating $\langle\hat{\phi}^2\rangle_0$}
\label{sec:cal}
We now set out to find an expression for the unrenormalized variance $\langle\hat{\phi}^2\rangle_0$ as a function of $\theta$. For this we follow \cite{Herranen:2013raa,Serreau:2011fu} closely, where we refer the reader for more details. The reader not interested in the details of the derivation may skip directly to the results, which are given in section \ref{sec:res}.
  
Since our ansatz for the full interacting mode leads to the same equation as in the free case, (\ref{eq:modehar}) with (\ref{eq:m2pi}) is (\ref{eq:modeireh}) with $\xi\rightarrow\theta$, the full interacting mode will be given by (\ref{eq:rehmode}) but where the index of the Hankel function is given by $\nu$ which is related to the  to-be-solved $\theta$
\ee{F_k(\eta)=\f{1}{2}\sqrt{-\pi\eta}H^{(1)}_\nu(-k\eta)\,;\qquad\nu^2=\f{1}{4}+12\bigg(\f{1}{6}-\theta\bigg)\f{1-3w}{(3w+1)^2}\,.\label{eq:nu3}}
We can then write the variance as 
\ee{\langle\hat{\phi}^2\rangle_0= \f{|\eta|}{8\pi a^2}\int^\infty_0 dk\,k^2\big|H^{(1)}_\nu (-k\eta)\big|^2\,.\label{eq:ints}}
In the situation where 
\ee{0<3/2-\nu\equiv \delta \ll 1\,,\label{eq:constr}}
i.e. when $\theta$ is such that $\nu$ is slightly smaller than $3/2$ the integral in (\ref{eq:ints}) can be solved analytically as was shown in \cite{Boyanovsky:2005sh,Serreau:2011fu} by using a cut-off regularization, for dimensional regularization see \cite{Herranen:2013raa}. From now on we will only consider situations where (\ref{eq:constr}) is satisfied and this is required for the validity of the derivation in \cite{Boyanovsky:2005sh,Serreau:2011fu}. Situations where this restriction cannot be imposed can also be approached analytically, see for example the works \cite{Glavan:2017jye,Glavan:2014uga,Janssen:2009pb}

Following \cite{Serreau:2011fu} we will first introduce a physical UV cut-off as
\ee{H|\eta|k\leq \Lambda_\text{\tiny UV}\quad \Rightarrow\quad k_\text{\tiny UV}=\Lambda_\text{\tiny UV}/(H|\eta|)\,,}
and furthermore in a similar fashion introduce the IR and "intermediate" (IM) cut-offs satisfying $\Lambda_\text{\tiny IR}\ll\Lambda_\text{\tiny IM}\ll\Lambda_\text{\tiny UV}$.

The integral (\ref{eq:ints}) may now be solved in a piecewise manner. For the UV contribution we can use the asymptotic expansion,
\ee{\big|H^{(1)}_\nu (x)\big|^2\overset{\rm UV}{\longrightarrow}\f{2}{\pi |x|}\bigg[1+\f{\nu^2-1/4}{2x^2}+\mathcal{O}(x^{-4})\bigg]\,,\label{eq:asyuv}}
which we note was extended for negative arguments by again using the relation $H^{(1)}_\nu (-|x|)$ $=-e^{-i\nu\pi}H^{(2)}_\nu (|x|)$. With the above and the formulae of the previous section we can write
\ea{\f{|\eta|}{8\pi a^2}\int^{k_\text{\tiny UV}}_{k_\text{\tiny IM}} dk\,k^2\big|H^{(1)}_\nu (-k\eta)\big|^2&=\f{1}{8\pi^2}\bigg\{\bigg(\f{\Lambda_{\text{\tiny UV}}}{Ha\eta}\bigg)^2+{R}{}\bigg(\f{1}{6}-\theta\bigg)\log\f{\Lambda_{\text{\tiny UV}}}{H}+\mathcal{O}(\Lambda_{\text{\tiny UV}}^{-1})\bigg\}\nonumber \\&-\f{1}{8\pi^2}\bigg\{\bigg(\f{\Lambda_{\text{\tiny IM}}}{Ha\eta}\bigg)^2+\f{2}{(a\eta )^{2}}\log\f{\Lambda_{\text{\tiny IM}}}{H}+\mathcal{O}(\delta,\Lambda_{\text{\tiny IM}}^{-1})\bigg\}\,.\label{eq:UV}}
Using very similar steps as for the UV contribution, for the middle piece we can use
\ee{\big|H^{(1)}_\nu (x)\big|^2\overset{\rm IM}{\longrightarrow}\big|H^{(1)}_{3/2} (x)\big|^2=\f{2}{\pi |x|^3}\big(1+x^2\big)\,,}
giving
\ee{\f{|\eta|}{8\pi a^2}\int^{k_\text{\tiny IM}}_{k_\text{\tiny IR}} dk\,k^2\big|H^{(1)}_\nu (-k\eta)\big|^2=\f{1}{8\pi^2}\bigg\{\bigg(\f{\Lambda_{\text{\tiny IM}}}{Ha\eta}\bigg)^2+\f{2}{(a\eta )^{2}}\log\f{\Lambda_{\text{\tiny IM}}}{\Lambda_{\text{\tiny IR}}}+\mathcal{O}(\delta,\Lambda_{\text{\tiny IR}})\bigg\}\,.\label{eq:IM}}
Finally, for the IR contribution we can use the familiar small argument form
\ee{|H^{(1)}_\nu(x)|^2\overset{\rm IR}{\longrightarrow}\big(2^\nu\Gamma[\nu]/(\pi |x|^\nu)\big)^2\,,\label{eq:asymp}}
to write
\ee{\f{|\eta|}{8\pi a^2}\int^{k_\text{\tiny IR}}_{0} dk\,k^2\big|H^{(1)}_\nu (-k\eta)\big|^2=\f{1}{8\pi^2(a\eta)^2}\bigg\{\f{1}{\delta}-2(2-\gamma_e-\log 2)+{2}\log\f{\Lambda_{\text{\tiny IR}}}{H}+\mathcal{O}(\delta,\Lambda_{\text{\tiny IR}})\bigg\}\,.\label{eq:IR}}
Combining (\ref{eq:UV}), (\ref{eq:IM}) and (\ref{eq:IR}) we can up to small terms write for the variance
\ee{\langle\hat{\phi}^2\rangle_0=\f{1}{8\pi^2}\bigg\{\bigg(\f{\Lambda_{\text{\tiny UV}}}{Ha\eta}\bigg)^2+R\bigg(\f{1}{6}-\theta\bigg)\log\f{\Lambda_{\text{\tiny UV}}'}{H}+\f{(a\eta )^{-2}}{\delta}\bigg\}+\cdots\,,\label{eq:vardiv}}
where we have absorbed the numerical factors $2(-2+\gamma_e+\log 2)$ in (\ref{eq:IR}) to a redefinition $\Lambda_{\text{\tiny UV}}\rightarrow \Lambda_{\text{\tiny UV}}'$ inside the logarithm, which is correct up to small terms that are beyond our approximation\footnote{To show this one may use results from section \ref{sec:mod} to write \ea{\nonumber(a\eta )^{2}R(1/6-\theta)\log\big( \Lambda e^{N/2}\big)&=(\nu^2-1/4)\log\big( \Lambda e^{N/2}\big)=\big(2-3\delta+\mathcal{O}(\delta^2)\big)\log\big( \Lambda e^{N/2}\big)\\ &=(a\eta )^{2}R(1/6-\theta)\log \Lambda +N +\mathcal{O}(\delta)\,,\nonumber} where $N$ is an arbitrary $\mathcal{O}(1)$ number.}.

To conclude this subsection, we point out that in addition to our approach, which follows the works \cite{Boyanovsky:2005sh,Serreau:2011fu} closely, it is possible to give the integral in (\ref{eq:ints}) an expression in closed from without relying on approximations that make use of an expansion in $\delta$ and involve splitting the integral in three pieces. This rather formal expression can be found for example in the seminal works \cite{BD,Chernikov:1968zm}.
\subsection{Renormalization}
Here we renormalize the variance obtained in the previous section in (\ref{eq:vardiv}). First we will split the result into a divergent piece and a finite physical contribution and from now on we drop the subscript "UV" from the cut-off
\ea{\langle\hat{\phi}^2\rangle_0&=\f{1}{8\pi^2}\bigg\{\bigg(\f{2\Lambda}{3w+1}\bigg)^2+R\bigg(\f{1}{6}-\theta\bigg)\bigg[\log\f{\Lambda'}{\mu}+\log\f{\mu}{H}\bigg]+\f{(a\eta )^{-2}}{\delta}\bigg\}\nonumber\\ &\equiv\f{1}{8\pi^2}\bigg\{\bigg(\f{2\Lambda}{3w+1}\bigg)^2+R\bigg(\f{1}{6}-\theta\bigg)\log\f{\Lambda'}{\mu}\bigg\} +\mathcal{F}
\,,\label{eq:m2pia}}
where $\mathcal{F}$ represents all the finite terms contained in $\langle\hat{\phi}^2\rangle_0$. With (\ref{eq:m2pia}) the gap equation (\ref{eq:m2pi}) becomes
\ea{\theta R
&=\xi R+3\lambda\mathcal{F}
+\bigg\{\delta m^2+\delta\xi R+3(\lambda+\delta\lambda)\f{\big(\f{2\Lambda}{3w+1}\big)^2+R\left(1/6-\theta\right)\log\big(\f{\Lambda'}{\mu}\big)}{8\pi^2}+3\delta\lambda\mathcal{F}
\bigg\}\,,\label{eq:m2pib}}
where the counter terms were introduced by the bare constants as $c_0\rightarrow c+\delta c$ and note that $m=0$. As our renormalization prescription we impose that the divergent expression in the wavy brackets of (\ref{eq:m2pib}) vanishes, 
from which follow the standard counter terms in the 2PI Hartree approximation \cite{Herranen:2013raa}
\ea{\delta\lambda&=\f{3\lambda^2\log\left(\Lambda'/\mu\right)}{8\pi^2}\bigg(1-\f{3\lambda\log\left(\Lambda'/\mu\right)}{8\pi^2}\bigg)^{-1}\nonumber\,,\\\delta \xi&=\f{3\lambda (\xi-1/6)\log\left(\Lambda'/\mu\right)}{8\pi^2}\bigg(1-\f{3\lambda\log\left(\Lambda'/\mu\right)}{8\pi^2}\bigg)^{-1} \nonumber\,, \\\delta m^2&=-\f{3\lambda \big(\f{2\Lambda}{3w+1}\big)^2}{8\pi^2}\bigg(1-\f{3\lambda\log\left(\Lambda'/\mu\right)}{8\pi^2}\bigg)^{-1}\,.}
With the divergences properly removed from $\langle\hat{\phi}^2\rangle_0$ the gap equation (\ref{eq:m2pi}) becomes a simple algebraic relation
\ee{\theta R=\tilde{\xi}R+\f{3\tilde{\lambda}}{8\pi^2}\f{(a\eta)^{-2}}{\delta}\,,\label{eq:gap2}}
where we have made the definitions
\ee{\tilde{\lambda}=\f{\lambda}{1-\f{3\lambda}{8\pi^2}\log\left(H/\mu\right)}\,,\qquad\tilde{\xi}=\f{1}{6}+\f{\xi-\f{1}{6}}{1-\f{3\lambda}{8\pi^2}\log\left(H/\mu\right)}\label{eq:runs}\,.}
As seen from (\ref{eq:runs}) the 2PI Hartree approximation as a by product results in running of the constants very similarly to what is found via 1-loop renormalization group improvement \cite{Pilaftsis:2017enx}. 

To conclude this section, we address an important issue concerning the form of the solution and the ansatz used. In our parametrization for the solutions in (\ref{eq:m2pi}) we assumed the $\theta$-parameter to be a constant and due to the presence of the running logarithms in (\ref{eq:runs}) this is strictly true only for de Sitter space. However, also for $w\neq-1$ this is a good approximation since the time dependence in $\theta$ is solely due to running, which is only logarithmic and thus quite mild. Furthermore, it is multiplied by the 1-loop suppression factor $\propto3\lambda/(8\pi^2)$. Because of this in what follows we will neglect the running in our results and make the replacements
\ee{\tilde{\lambda}~~\rightarrow~~\lambda\,,\qquad\tilde{\xi}~~\rightarrow~~\xi\,.}

With the results of this section we can write down a form for the gap equation that can be solved for an arbitrary equation of state, with which we may investigate the generation of a non-zero variance, $\langle\hat{\phi}^2\rangle$.

\section{Results}
\label{sec:res}
Before addressing the results in several important cases, we provide a simple picture allowing one to intuitively understand why a particular background leads to an excitation of the field. 

In the renormalized form in the 2PI Hartree approximation the mode equation for a massless, light and self-interacting scalar field reads
\beq
F''_{k}(\eta)+ \bigg\{\bk^2 + a^2  \bigg[\left(\xi-\frac{1}{6}\right) R +3\lambda\langle\hat{\phi}^2\rangle\bigg]\bigg\} F_{k}(\eta) = 0 \,.\label{eq:modehar2}
\eeq
Considering the initial condition where $\langle\hat{\phi}^2\rangle$ starts out as small from the above it is easy to see that for many combinations of $\xi$ and the equation of state parameter $w$ the expression inside the square brackets leads the IR quantum modes to obtain a negative effective mass squared due to the presence of the background curvature. Broadly speaking, this is an unstable configuration akin to tachyonic or spinodal instability \cite{Felder:2001kt,Felder:2000hj} and can be expected to result in strong amplification of the field. Since $R=3(1-3w)H^2$ we can conclude that for $w<1/3$ amplification occurs when $\xi\lesssim1/6$, and similarly for $w>1/3$ when $\xi \gtrsim 1/6$. One may also see from (\ref{eq:modehar2}) that the more the field is amplified the more the growing variance $\langle\hat{\phi}^2\rangle$ will via interactions work against further excitation. Finally, an equilibrium between the two competing effects will be reached which is precisely the value solved by our approach. Based on this argument the magnitude of the variance is expected to saturate close to the point where the IR divergence disappears, or more descriptively becomes self regulated, since at this point the field is no longer light due to the effective mass generated by interactions $\propto\lambda\langle\hat{\phi}^2\rangle$. From equation (\ref{eq:freev}) one may see that this happens when the index in the mode function is smaller than $3/2$, which is the reason we can expect an expansion in $3/2-\nu$ to be valid in cases where there is IR amplification.

For completeness, we start by laying out the steps one needs to make for finding an expression for $\langle\hat{\phi}^2\rangle$. Neglecting the small effect from the running of couplings, the gap equation (\ref{eq:gap2}) reads
\ee{\theta R={\xi}R+\f{3{\lambda}}{8\pi^2}\f{(a\eta)^{-2}}{\delta}\,,\label{eq:gap3}}
and once we have solved from it an expression for $\theta$, the variance comes from the renormalized form of (\ref{eq:m2pi})
\ee{\langle\hat{\phi}^2\rangle=\f{\theta-\xi}{3\lambda} R\,.\label{eq:varri}}
The $a\eta$ and $R$ for a general equation of state $w$ are given in (\ref{eq:rehaH2}) and (\ref{eq:R}), respectively. Importantly, our main focus lies in cases where an IR amplification takes place and the above provide valid solutions only in such a case: the $\delta$ in (\ref{eq:gap3}) is required to be a small positive parameter and it is related to $\nu$ as given in (\ref{eq:nu3}) by $\delta = 3/2-\nu$. Mathematically, $\delta$ is required to be small so that the expansion (and truncation) in $\delta$ used in the derivation in section \ref{sec:cal} is valid and positive so that a finite IR limit for the variance exists.

\subsection{Examples}
Now we apply our formalism to several cosmologically relevant cases.
\subsection*{$w=-1,$ $~\xi=\,0$}
It is already well-established  that a light, minimally coupled scalar field with a quartic self-interaction will it in de Sitter space exhibit random walk and as a result generate a non-zero variance. This case is often investigated via the stochastic approach, which is able to probe also the non-Gaussian features of the distribution unlike the 2PI Hartree approximation as used here \cite{Starobinsky:1994bd}. Nonetheless, the $w=-1$ case provides a useful check of our results.

The gap equation (\ref{eq:gap3}) in de Sitter space with $\xi =0$ gives to leading order in $\delta$
\ee{12H^2\theta=\f{3\lambda}{8\pi^2}\f{H^2}{4\theta}\qquad\Leftrightarrow\qquad\theta=\f{1}{4}\sqrt{\f{\lambda}{8\pi^2}}\,,}
which leads to the variance via (\ref{eq:varri})
\ee{\langle\hat{\phi}^2\rangle=\f{H^2}{\pi\sqrt{8\lambda}}\,,}
which is the well-known result derived for example in \cite{Herranen:2013raa,Starobinsky:1994bd}.
\subsection*{$w=-1,$ $~\xi\gg1$}
It is also instructive to study how our formalism fares in a case where there obviously is no IR enhancement and no variance is generated. This happens when the field is effectively heavy, which in de Sitter space may be achieved by having a large non-minimal coupling. In this scenario the gap equation becomes
\ee{12H^2\theta=12H^2\xi+\f{3\lambda}{8\pi^2}\f{H^2}{4\theta}
\,.}
It is simple to show that for $\xi\gg1$ the above has two solutions for $\theta$ neither of which giving $1\gg3/2-\nu>0$ and thus satisfying the required conditions as discussed in the beginning of this section. This indicates that for this choice a solution resulting in IR enhancement does not exist, 
as expected.

Next we can turn to cases which to the best of our knowledge have not previously been addressed in literature.
 
\subsection*{$w=0,$ $~\xi=\,0$}
The equation of state $w=0$ is frequently encountered in cosmology, for example for a matter dominated Universe or often during reheating after inflation when the inflaton oscillates around its minimum. Like in de Sitter space for a minimally coupled theory equation (\ref{eq:modehar2}) has a negative contribution from $-R/6$, which is expected to lead to $\langle\hat{\phi}^2\rangle>0$ via tachyonic amplification. The gap equation is, again to leading order in $\delta$
\ee{3H^2\theta=\f{3\lambda}{8\pi^2}\f{H^2/4}{4\theta}\qquad\Leftrightarrow\qquad\theta=\f{1}{4}\sqrt{\f{\lambda}{8\pi^2}}\,,}
and the variance reads
\ee{\langle\hat{\phi}^2\rangle=\f{H^2}{4\pi\sqrt{8\lambda}}\,.}
So very much like on a de Sitter background, for $w=0$ a minimally coupled light field will become excited by the expansion of space.
\subsection*{$w=1,$ $~\xi\gg1$}
An equation of state with $w=1$ is usually called kination and is encountered for example in cases where the potential of the inflaton becomes negligibly small after inflation has ended. Since now $R<0$, we expect a non-zero variance to be generated only for large $\xi$.
It proves convenient to define a new variable $\theta'$ as $\nu^2=9/4(1-\theta')$, with which one gets to leading order
\ea{-9\bigg[1+\mathcal{O}(\xi^{-1})\bigg]H^2=-9H^2+\bigg[9-6\xi + \f{2\lambda}{\pi^2\theta'}\bigg]H^2\quad \Leftrightarrow\quad \theta'=
\f{2\lambda}{\pi^2(6\xi-9) }+\mathcal{O}(\xi^{-3})\,,}
resulting at the limit of a large $\xi$ to
\ee{\langle\hat{\phi}^2\rangle=\f{2\xi}{\lambda}H^2\,,}
providing yet another example where the background curvature can induce a non-zero variance.
\subsection*{$w=-1/3,$ $~\xi=0$}
The so-called Milne Universe can be described with $w=-1/3$, which is also the point after which, if inflation took place, the co-moving Hubble horizon starts increasing in size. This is a special point in the solutions as here conformal time (\ref{eq:rehaH2}) and the index of the mode function (\ref{eq:ints}) have singular behaviour. But since we have well-defined solutions when $w\neq-1/3$, above and below, and furthermore since the scalar curvature is perfectly finite at this point, the pole is not visible in the physical results, for example for the variance.

Assuming that we are arbitrary close to the divergent point, $w=-1/3\pm\epsilon$, it again helps to define a new variable, $\nu^2=9/4(1-\theta')$, with which we get
\ea{\bigg[1+\mathcal{O}(\epsilon^2)\bigg]H^2=\f{9\lambda}{8\pi^2\theta'}(\epsilon H)^2\quad \Leftrightarrow\quad \theta'
=\frac{9 \lambda}{8 \pi ^2} \epsilon ^2+\mathcal{O}(\epsilon^4)\,,}
which when setting $\epsilon\rightarrow0$ gives
\ee{\langle\hat{\phi}^2\rangle=\f{H^2}{3\lambda}\,,}
Showing that indeed a finite result for a non-zero variance is also present for $w=-1/3$, as anticipated.
\section{Discussion}
In this work we have shown that a generic effect for a light scalar field $\phi$ on a curved background is the generation of a non-zero variance $\langle\hat{\phi}^2\rangle$. Our model contained quartic self-interactions that were analysed non-perturbatively, specifically by using the leading order truncation in the 2-particle-irreducible (2PI) Feynman diagram expansion, also known as the Hartree approximation. The background was assumed to be characterized by the cosmological equation of state parameter $w$ relating the energy and pressure densities as $p=w\rho $.

A natural direction beyond the results presented here is relaxing the requirement of the field $\phi$ being simply a spectator on a curved background. In the situation where it also results in a non-trivial gravitational backreaction it seems plausible that one may no longer neglect the quantum nature of gravity and fluctuations of the metric must be included. In particular, equations of state with $w>-1/3$ seem promising as it is well-known in the context of Early Universe cosmology that in quasi de Sitter spaces the difference between semi-classical gravity and quantising the metric fluctuations amounts to different factors of the slow-roll $\epsilon$ parameter, which is small for $w<-1/3$ but order unity for $w>-1/3$.

There are several obvious and interesting applications for our results. Gravitational amplification of scalars during an epoch with $w=0$ has direct implications for the reheating of the Universe after inflation. Furthermore, a gravitationally generated effective mass can, if large enough, have an impact on the parametric resonance mechanism that potentially takes place immediately after the end of inflation, or in short preheating \cite{Kofman}.

Another application worth exploring is the behaviour of the Higgs field in the Early Universe, in particular the possible destabilization of the electroweak vacuum \cite{Herranen:2014cua,Herranen:2015ima}. Previously, only $w=-1$ and $w=0$ have been investigated, however our results indicate that in all cosmological epochs besides radiation domination the Higgs field can become excited, potentially enough to trigger vacuum decay.

Perhaps the most promising application of this work is a novel framework for dark matter (DM) generation. In itself, gravitational dark matter generation is of course not a new idea, and for example the WIMPzilla models have received significant attention and are currently actively studied \cite{KolbLong}. Furthermore, extremely potent dark matter production via tachyonic resonance induced by the oscillating scalar curvature during reheating was discovered in \cite{gravdm} to allow purely gravitational DM production with the correct abundance also for very light particles. However, our formalism making use of the simple parametrization via the cosmological equation of state $w$ provides previously unknown generic solutions for a variety of situations that require a thorough investigation. 
\acknowledgments{TM would like to thank Arttu Rajantie, Konstantinos Dimopoulos, Gerasimos Rigopoulos and Sami Nurmi for illuminating discussions and Julien Serreau for valuable comments on the finished manuscript. The research leading to these results has received funding from the European Research Council under the European Union's Horizon 2020 program (ERC Grant Agreement no.648680) and from the STFC grant ST/P000762/1. }


\begin{thebibliography}{99}

\bibitem{Birrell:1982ix}
  N.~D.~Birrell and P.~C.~W.~Davies,
  ``Quantum Fields in Curved Space,''
  doi:10.1017/CBO9780511622632

\bibitem{ParkerToms}
L. Parker and D. J. Toms
"Quantum Field Theory in Curved Space-time: Quantized Fields and Gravity"
Cambridge University Press, 2009, 500 p

\bibitem{Tsamis:2005hd}
  N.~C.~Tsamis and R.~P.~Woodard,
  ``Stochastic quantum gravitational inflation,''
  Nucl.\ Phys.\ B {\bf 724} (2005) 295
  doi:10.1016/j.nuclphysb.2005.06.031
  [gr-qc/0505115].
  
  
\bibitem{Kahya:2010xh}
  E.~O.~Kahya, V.~K.~Onemli and R.~P.~Woodard,
  ``The Zeta-Zeta Correlator Is Time Dependent,''
  Phys.\ Lett.\ B {\bf 694} (2011) 101
  doi:10.1016/j.physletb.2010.09.050
  [arXiv:1006.3999 [astro-ph.CO]].


\bibitem{Seery:2010kh}
  D.~Seery,
  ``Infrared effects in inflationary correlation functions,''
  Class.\ Quant.\ Grav.\  {\bf 27} (2010) 124005
  doi:10.1088/0264-9381/27/12/124005
  [arXiv:1005.1649 [astro-ph.CO]].

\bibitem{Enqvist:2008kt}
  K.~Enqvist, S.~Nurmi, D.~Podolsky and G.~I.~Rigopoulos,
  ``On the divergences of inflationary superhorizon perturbations,''
  JCAP {\bf 0804} (2008) 025
  doi:10.1088/1475-7516/2008/04/025
  [arXiv:0802.0395 [astro-ph]].

\bibitem{Burgess:2009bs}
  C.~P.~Burgess, L.~Leblond, R.~Holman and S.~Shandera,
  ``Super-Hubble de Sitter Fluctuations and the Dynamical RG,''
  JCAP {\bf 1003} (2010) 033
  doi:10.1088/1475-7516/2010/03/033
  [arXiv:0912.1608 [hep-th]].
  
\bibitem{Starobinsky:1994bd}
  A.~A.~Starobinsky and J.~Yokoyama,
  ``Equilibrium state of a selfinteracting scalar field in the De Sitter background,''
  Phys.\ Rev.\ D {\bf 50} (1994) 6357
  doi:10.1103/PhysRevD.50.6357
  [astro-ph/9407016].
  
\bibitem{Garbrecht:2014dca}
  B.~Garbrecht, F.~Gautier, G.~Rigopoulos and Y.~Zhu,
  ``Feynman Diagrams for Stochastic Inflation and Quantum Field Theory in de Sitter Space,''
  Phys.\ Rev.\ D {\bf 91} (2015) 063520
  doi:10.1103/PhysRevD.91.063520
  [arXiv:1412.4893 [hep-th]].
  
\bibitem{Garbrecht:2013coa}
  B.~Garbrecht, G.~Rigopoulos and Y.~Zhu,
  ``Infrared correlations in de Sitter space: Field theoretic versus stochastic approach,''
  Phys.\ Rev.\ D {\bf 89} (2014) 063506
  doi:10.1103/PhysRevD.89.063506
  [arXiv:1310.0367 [hep-th]].
  
\bibitem{Garbrecht:2011gu}
  B.~Garbrecht and G.~Rigopoulos,
  ``Self Regulation of Infrared Correlations for Massless Scalar Fields during Inflation,''
  Phys.\ Rev.\ D {\bf 84} (2011) 063516
  doi:10.1103/PhysRevD.84.063516
  [arXiv:1105.0418 [hep-th]].
  

\bibitem{Serreau:2011fu}
  J.~Serreau,
  ``Effective potential for quantum scalar fields on a de Sitter geometry,''
  Phys.\ Rev.\ Lett.\  {\bf 107} (2011) 191103
  doi:10.1103/PhysRevLett.107.191103
  [arXiv:1105.4539 [hep-th]].

\bibitem{Starobinsky:1986fx}
  A.~A.~Starobinsky,
  ``Stochastic De Sitter (inflationary) Stage In The Early Universe,''
  Lect.\ Notes Phys.\  {\bf 246} (1986) 107.
  doi:10.1007/3-540-16452-9 6

\bibitem{Nakao:1988yi} 
  K.~i.~Nakao, Y.~Nambu and M.~Sasaki,
  ``Stochastic Dynamics of New Inflation,''
  Prog.\ Theor.\ Phys.\  {\bf 80}, 1041 (1988).
  doi:10.1143/PTP.80.1041
  
\bibitem{Nambu:1988je} 
  Y.~Nambu and M.~Sasaki,
  ``Stochastic Approach to Chaotic Inflation and the Distribution of Universes,''
  Phys.\ Lett.\ B {\bf 219}, 240 (1989).
  doi:10.1016/0370-2693(89)90385-7
  
\bibitem{Linde:1993xx}
  A.~D.~Linde, D.~A.~Linde and A.~Mezhlumian,
  ``From the Big Bang theory to the theory of a stationary universe,''
  Phys.\ Rev.\ D {\bf 49} (1994) 1783
  doi:10.1103/PhysRevD.49.1783
  [gr-qc/9306035].

\bibitem{Prokopec:2007ak}
  T.~Prokopec, N.~C.~Tsamis and R.~P.~Woodard,
  ``Stochastic Inflationary Scalar Electrodynamics,''
  Annals Phys.\  {\bf 323} (2008) 1324
  doi:10.1016/j.aop.2007.08.008
  [arXiv:0707.0847 [gr-qc]].

\bibitem{Finelli:2008zg} 
  F.~Finelli, G.~Marozzi, A.~A.~Starobinsky, G.~P.~Vacca and G.~Venturi,
  ``Generation of fluctuations during inflation: Comparison of stochastic and field-theoretic approaches,''
  Phys.\ Rev.\ D {\bf 79}, 044007 (2009)
  doi:10.1103/PhysRevD.79.044007
  [arXiv:0808.1786 [hep-th]].
  
  
\bibitem{Finelli:2010sh} 
  F.~Finelli, G.~Marozzi, A.~A.~Starobinsky, G.~P.~Vacca and G.~Venturi,
  ``Stochastic growth of quantum fluctuations during slow-roll inflation,''
  Phys.\ Rev.\ D {\bf 82}, 064020 (2010)
  doi:10.1103/PhysRevD.82.064020
  [arXiv:1003.1327 [hep-th]].

\bibitem{Prokopec:2011ms}
  T.~Prokopec,
  ``Symmetry breaking and the Goldstone theorem in de Sitter space,''
  JCAP {\bf 1212} (2012) 023
  doi:10.1088/1475-7516/2012/12/023
  [arXiv:1110.3187 [gr-qc]].

 \bibitem{Kainulainen:2016vzv}
  K.~Kainulainen, S.~Nurmi, T.~Tenkanen, K.~Tuominen and V.~Vaskonen,
  ``Isocurvature Constraints on Portal Couplings,''
  JCAP {\bf 1606} (2016) no.06,  022
  doi:10.1088/1475-7516/2016/06/022
  [arXiv:1601.07733 [astro-ph.CO]].
    
\bibitem{Vennin:2015hra} 
  V.~Vennin and A.~A.~Starobinsky,
  ``Correlation Functions in Stochastic Inflation,''
  Eur.\ Phys.\ J.\ C {\bf 75}, 413 (2015)
  doi:10.1140/epjc/s10052-015-3643-y
  [arXiv:1506.04732 [hep-th]].
    
\bibitem{Nurmi:2015ema}
  S.~Nurmi, T.~Tenkanen and K.~Tuominen,
  ``Inflationary Imprints on Dark Matter,''
  JCAP {\bf 1511} (2015) no.11,  001
  doi:10.1088/1475-7516/2015/11/001
  [arXiv:1506.04048 [astro-ph.CO]].
    
\bibitem{Assadullahi:2016gkk}
  H.~Assadullahi, H.~Firouzjahi, M.~Noorbala, V.~Vennin and D.~Wands,
  ``Multiple Fields in Stochastic Inflation,''
  JCAP {\bf 1606} (2016) no.06,  043
  doi:10.1088/1475-7516/2016/06/043
  [arXiv:1604.04502 [hep-th]].
  
  
\bibitem{Vennin:2016wnk}
  V.~Vennin, H.~Assadullahi, H.~Firouzjahi, M.~Noorbala and D.~Wands,
  ``Critical Number of Fields in Stochastic Inflation,''
  Phys.\ Rev.\ Lett.\  {\bf 118} (2017) no.3,  031301
  doi:10.1103/PhysRevLett.118.031301
  [arXiv:1604.06017 [astro-ph.CO]].
  
\bibitem{Moss:2016uix}
  I.~Moss and G.~Rigopoulos,
  ``Effective long wavelength scalar dynamics in de Sitter,''
  JCAP {\bf 1705} (2017) no.05,  009
  doi:10.1088/1475-7516/2017/05/009
  [arXiv:1611.07589 [gr-qc]].
  
\bibitem{Hardwick:2017fjo}
  R.~J.~Hardwick, V.~Vennin, C.~T.~Byrnes, J.~Torrado and D.~Wands,
  ``The stochastic spectator,''
  arXiv:1701.06473 [astro-ph.CO].
      
\bibitem{Karakaya:2017evp}
  G.~Karakaya and V.~K.~Onemli,
  ``Quantum Effects of Mass on Scalar Field Correlations and Fluctuations during Inflation,''
  arXiv:1710.06768 [gr-qc].
    
\bibitem{Cornwall:1974vz}
  J.~M.~Cornwall, R.~Jackiw and E.~Tomboulis,
  ``Effective Action for Composite Operators,''
  Phys.\ Rev.\ D {\bf 10} (1974) 2428.
  doi:10.1103/PhysRevD.10.2428
  
\bibitem{Berges:2004yj}
  J.~Berges,
  ``Introduction to nonequilibrium quantum field theory,''
  AIP Conf.\ Proc.\  {\bf 739} (2005) 3
  doi:10.1063/1.1843591
  [hep-ph/0409233].
  
\bibitem{Blaizot:2001nr}
  J.~P.~Blaizot and E.~Iancu,
  ``The Quark gluon plasma: Collective dynamics and hard thermal loops,''
  Phys.\ Rept.\  {\bf 359} (2002) 355
  doi:10.1016/S0370-1573(01)00061-8
  [hep-ph/0101103].
    
\bibitem{Ramsey:1997qc}
  S.~A.~Ramsey and B.~L.~Hu,
  ``O(N) quantum fields in curved space-time,''
  Phys.\ Rev.\ D {\bf 56} (1997) 661
  doi:10.1103/PhysRevD.56.661
  [gr-qc/9706001].

\bibitem{Sloth:2006az}
  M.~S.~Sloth,
  ``On the one loop corrections to inflation and the CMB anisotropies,''
  Nucl.\ Phys.\ B {\bf 748} (2006) 149
  doi:10.1016/j.nuclphysb.2006.04.029
  [astro-ph/0604488].

\bibitem{Riotto:2008mv}
  A.~Riotto and M.~S.~Sloth,
  ``On Resumming Inflationary Perturbations beyond One-loop,''
  JCAP {\bf 0804} (2008) 030
  doi:10.1088/1475-7516/2008/04/030
  [arXiv:0801.1845 [hep-ph]].
  
\bibitem{Tranberg:2008ae}
  A.~Tranberg,
  ``Quantum field thermalization in expanding backgrounds,''
  JHEP {\bf 0811} (2008) 037
  doi:10.1088/1126-6708/2008/11/037
  [arXiv:0806.3158 [hep-ph]].
  
\bibitem{Arai:2011dd}
  T.~Arai,
  ``Nonperturbative Infrared Effects for Light Scalar Fields in de Sitter Space,''
  Class.\ Quant.\ Grav.\  {\bf 29} (2012) 215014
  doi:10.1088/0264-9381/29/21/215014
  [arXiv:1111.6754 [hep-th]].
  
\bibitem{Arai:2012sh}
  T.~Arai,
  ``Renormalization of the 2PI Hartree-Fock approximation on de Sitter background in the broken phase,''
  Phys.\ Rev.\ D {\bf 86} (2012) 104064
  doi:10.1103/PhysRevD.86.104064
  [arXiv:1204.0476 [hep-th]].

 
\bibitem{Serreau:2013psa}
  J.~Serreau and R.~Parentani,
  ``Nonperturbative resummation of de Sitter infrared logarithms in the large-N limit,''
  Phys.\ Rev.\ D {\bf 87} (2013) 085012
  doi:10.1103/PhysRevD.87.085012
  [arXiv:1302.3262 [hep-th]].


\bibitem{Nacir:2013xca}
  D.~L.~Lopez Nacir, F.~D.~Mazzitelli and L.~G.~Trombetta,
  ``Hartree approximation in curved spacetimes revisited: The effective potential in de Sitter spacetime,''
  Phys.\ Rev.\ D {\bf 89} (2014) 024006
  doi:10.1103/PhysRevD.89.024006
  [arXiv:1309.0864 [hep-th]].
  

\bibitem{Herranen:2013raa}
  M.~Herranen, T.~Markkanen and A.~Tranberg,
  ``Quantum corrections to scalar field dynamics in a slow-roll space-time,''
  JHEP {\bf 1405} (2014) 026
  doi:10.1007/JHEP05(2014)026
  [arXiv:1311.5532 [hep-ph]].
  
\bibitem{Ziaeepour:2017wdp}
  H.~Ziaeepour,
  ``Non-equilibrium evolution of quantum fields during inflation and late accelerating expansion,''
  arXiv:1711.01925 [gr-qc].

\bibitem{Gautier:2013aoa}
  F.~Gautier and J.~Serreau,
  ``Infrared dynamics in de Sitter space from Schwinger-Dyson equations,''
  Phys.\ Lett.\ B {\bf 727} (2013) 541
  doi:10.1016/j.physletb.2013.10.072
  [arXiv:1305.5705 [hep-th]].

\bibitem{Serreau:2013eoa}
  J.~Serreau,
  ``Renormalization group flow and symmetry restoration in de Sitter space,''
  Phys.\ Lett.\ B {\bf 730} (2014) 271
  doi:10.1016/j.physletb.2014.01.058
  [arXiv:1306.3846 [hep-th]].
  
\bibitem{Guilleux:2015pma}
  M.~Guilleux and J.~Serreau,
  ``Quantum scalar fields in de Sitter space from the nonperturbative renormalization group,''
  Phys.\ Rev.\ D {\bf 92} (2015) no.8,  084010
  doi:10.1103/PhysRevD.92.084010
  [arXiv:1506.06183 [hep-th]].
  
\bibitem{Guilleux:2016oqv}
  M.~Guilleux and J.~Serreau,
  ``Nonperturbative renormalization group for scalar fields in de Sitter space: beyond the local potential approximation,''
  Phys.\ Rev.\ D {\bf 95} (2017) no.4,  045003
  doi:10.1103/PhysRevD.95.045003
  [arXiv:1611.08106 [gr-qc]].

\bibitem{Ford:1977in}
  L.~H.~Ford and L.~Parker,
  ``Infrared Divergences in a Class of Robertson-Walker Universes,''
  Phys.\ Rev.\ D {\bf 16} (1977) 245.
  doi:10.1103/PhysRevD.16.245
  
\bibitem{Lankinen:2016ile}
  J.~Lankinen and I.~Vilja,
  ``Gravitational Particle Creation in a Stiff Matter Dominated Universe,''
  JCAP {\bf 1708} (2017) 025
  doi:10.1088/1475-7516/2017/08/025
  [arXiv:1612.02586 [gr-qc]].
  

\bibitem{Lankinen:2017rvi}
  J.~Lankinen and I.~Vilja,
  ``Decay of a Massive Particle in a Stiff Matter Dominated Universe,''
  arXiv:1709.07236 [gr-qc].

\bibitem{Higuchi:2017sgj}
  A.~Higuchi and N.~Rendell,
  ``Infrared divergences for free quantum fields in cosmological spacetimes,''
  arXiv:1711.03964 [gr-qc].

\bibitem{Glavan:2013mra}
  D.~Glavan, T.~Prokopec and V.~Prymidis,
  ``Backreaction of a massless minimally coupled scalar field from inflationary quantum fluctuations,''
  Phys.\ Rev.\ D {\bf 89} (2014) no.2,  024024
  doi:10.1103/PhysRevD.89.024024
  [arXiv:1308.5954 [gr-qc]].

\bibitem{Cho:2015pwa}
  G.~Cho, C.~H.~Kim and H.~Kitamoto,
  ``Stochastic Dynamics of Infrared Fluctuations in Accelerating Universe,''
  doi:10.1142/9789813203952\_0018
  arXiv:1508.07877 [hep-th].

\bibitem{Prokopec:2015owa}
  T.~Prokopec,
  ``Late time solution for interacting scalar in accelerating spaces,''
  JCAP {\bf 1511} (2015) no.11,  016
  doi:10.1088/1475-7516/2015/11/016
  [arXiv:1508.07874 [gr-qc]].

\bibitem{Glavan:2015cut}
  D.~Glavan, T.~Prokopec and T.~Takahashi,
  ``Late-time quantum backreaction of a very light nonminimally coupled scalar,''
  Phys.\ Rev.\ D {\bf 94} (2016) 084053
  doi:10.1103/PhysRevD.94.084053
  [arXiv:1512.05329 [gr-qc]].

\bibitem{Glavan:2017jye}
  D.~Glavan, T.~Prokopec and A.~A.~Starobinsky,
  ``Stochastic dark energy from inflationary quantum fluctuations,''
  arXiv:1710.07824 [astro-ph.CO].

\bibitem{Misner:1974qy}
  C.~W.~Misner, K.~S.~Thorne and J.~A.~Wheeler,
  ``Gravitation,''
  San Francisco 1973, W. H. Freeman and Company, p. 1279.
    
\bibitem{Chernikov:1968zm}
  N.~A.~Chernikov and E.~A.~Tagirov,
  ``Quantum theory of scalar fields in de Sitter space-time,''
  Annales Poincare Phys.\ Theor.\ A {\bf 9} (1968) 109.
\bibitem{BD} 
  T.~S.~Bunch and P.~C.~W.~Davies,
  ``Quantum Field Theory in de Sitter Space: Renormalization by Point Splitting,''
  Proc.\ Roy.\ Soc.\ Lond.\ A {\bf 360} (1978) 117.


\bibitem{Markkanen:2016aes}
  T.~Markkanen and A.~Rajantie,
  ``Massive scalar field evolution in de Sitter,''
  JHEP {\bf 1701} (2017) 133
  doi:10.1007/JHEP01(2017)133
  [arXiv:1607.00334 [gr-qc]].
      
\bibitem{Kapusta:2006pm}
  J.~I.~Kapusta and C.~Gale,
  ``Finite-temperature field theory: Principles and applications,''

\bibitem{Allen:1985ux}
  B.~Allen,
  ``Vacuum States in de Sitter Space,''
  Phys.\ Rev.\ D {\bf 32} (1985) 3136.
  doi:10.1103/PhysRevD.32.3136
      
\bibitem{Cooper:1994hr}
  F.~Cooper, S.~Habib, Y.~Kluger, E.~Mottola, J.~P.~Paz and P.~R.~Anderson,
  ``Nonequilibrium quantum fields in the large N expansion,''
  Phys.\ Rev.\ D {\bf 50} (1994) 2848
  doi:10.1103/PhysRevD.50.2848
  [hep-ph/9405352].
  
      
  
\bibitem{Cooper:1996ii}
  F.~Cooper, S.~Habib, Y.~Kluger and E.~Mottola,
  ``Nonequilibrium dynamics of symmetry breaking in lambda Phi**4 field theory,''
  Phys.\ Rev.\ D {\bf 55} (1997) 6471
  doi:10.1103/PhysRevD.55.6471
  [hep-ph/9610345].

\bibitem{Boyanovsky:2005sh}
  D.~Boyanovsky, H.~J.~de Vega and N.~G.~Sanchez,
  ``Quantum corrections to slow roll inflation and new scaling of superhorizon fluctuations,''
  Nucl.\ Phys.\ B {\bf 747} (2006) 25
  doi:10.1016/j.nuclphysb.2006.04.010
  [astro-ph/0503669].

\bibitem{Glavan:2014uga}
  D.~Glavan, T.~Prokopec and D.~C.~van der Woude,
  Phys.\ Rev.\ D {\bf 91} (2015) no.2,  024014
  doi:10.1103/PhysRevD.91.024014
  [arXiv:1408.4705 [gr-qc]].

\bibitem{Janssen:2009pb}
  T.~M.~Janssen, S.~P.~Miao, T.~Prokopec and R.~P.~Woodard,
  JCAP {\bf 0905} (2009) 003
  doi:10.1088/1475-7516/2009/05/003
  [arXiv:0904.1151 [gr-qc]].
  
\bibitem{Pilaftsis:2017enx}
  A.~Pilaftsis and D.~Teresi,
  ``Exact RG Invariance and Symmetry Improved 2PI Effective Potential,''
  Nucl.\ Phys.\ B {\bf 920} (2017) 298
  doi:10.1016/j.nuclphysb.2017.04.015
  [arXiv:1703.02079 [hep-ph]].

\bibitem{Felder:2000hj}
  G.~N.~Felder, J.~Garcia-Bellido, P.~B.~Greene, L.~Kofman, A.~D.~Linde and I.~Tkachev,
  ``Dynamics of symmetry breaking and tachyonic preheating,''
  Phys.\ Rev.\ Lett.\  {\bf 87} (2001) 011601
  doi:10.1103/PhysRevLett.87.011601
  [hep-ph/0012142].

\bibitem{Felder:2001kt}
  G.~N.~Felder, L.~Kofman and A.~D.~Linde,
  ``Tachyonic instability and dynamics of spontaneous symmetry breaking,''
  Phys.\ Rev.\ D {\bf 64} (2001) 123517
  doi:10.1103/PhysRevD.64.123517
  [hep-th/0106179].

\bibitem{Kofman}
  L.~Kofman, A.~D.~Linde and A.~A.~Starobinsky,
  ``Towards the theory of reheating after inflation,''
  Phys.\ Rev.\ D {\bf 56} (1997) 3258
  doi:10.1103/PhysRevD.56.3258
  [hep-ph/9704452].   
  
\bibitem{Herranen:2014cua}
  M.~Herranen, T.~Markkanen, S.~Nurmi and A.~Rajantie,
  ``Spacetime curvature and the Higgs stability during inflation,''
  Phys.\ Rev.\ Lett.\  {\bf 113} (2014) no.21,  211102
  [arXiv:1407.3141 [hep-ph]].
 
\bibitem{Herranen:2015ima}
  M.~Herranen, T.~Markkanen, S.~Nurmi and A.~Rajantie,
  ``Spacetime curvature and Higgs stability after inflation,''
  Phys.\ Rev.\ Lett.\  {\bf 115} (2015) 241301
  [arXiv:1506.04065 [hep-ph]].
  
  \bibitem{KolbLong}
  R.~W.~Kolb and A.~J.~Long,
  ``Superheavy dark matter through Higgs portal operators,''
  Phys.\ Rev.\ D {\bf 96} (2017) no.10,  103540
  doi:10.1103/PhysRevD.96.103540
  [arXiv:1708.04293 [astro-ph.CO]].
  
  \bibitem{gravdm}
  T.~Markkanen and S.~Nurmi,
  ``Dark matter from gravitational particle production at reheating,''
  JCAP {\bf 1702} (2017) 008
  doi:10.1088/1475-7516/2017/02/008
  [arXiv:1512.07288 [astro-ph.CO]].
  
\end{thebibliography}
\end{document}